\documentclass[12pt,epsf,amstex]{article}
\usepackage [dvips]{graphicx}
\usepackage{amsmath}
\usepackage{amssymb}
\usepackage{epsfig}

\addtocounter{secnumdepth}{1}
\begin{document}

\newcommand{\br}{\bar{r}}
\newcommand{\bbeta}{\bar{\beta}}
\newcommand{\bgamma}{\bar{\gamma}}
\newcommand{\tbeta}{\tilde{\beta}}
\newcommand{\tgamma}{\tilde{\gamma}}
\newcommand{\bE}{{\bf{E}}}
\newcommand{\bO}{{\bf{O}}}
\newcommand{\bP}{{\bf{P}}}
\newcommand{\bR}{{\bf{R}}}
\newcommand{\bS}{{\bf{S}}}
\newcommand{\bT}{\mbox{\bf T}}
\newcommand{\bt}{\mbox{\bf t}}
\newcommand{\half}{\frac{1}{2}}
\newcommand{\thalf}{\tfrac{1}{2}}
\newcommand{\summ}{\sum_{m=1}^n}
\newcommand{\sumq}{\sum_{q=1}^\infty}
\newcommand{\sumqno}{\sum_{q\neq 0}}
\newcommand{\prodm}{\prod_{m=1}^n}
\newcommand{\prodq}{\prod_{q=1}^\infty}
\newcommand{\maxm}{\max_{1\leq m\leq n}}
\newcommand{\maxphi}{\max_{0\leq\phi\leq 2\pi}}
\newcommand{\tsum}{\Sigma}
\newcommand{\bsA}{\mathbf{A}}
\newcommand{\bsV}{\mathbf{V}}
\newcommand{\bsE}{\mathbf{E}}
\newcommand{\bsT}{\mathbf{T}}
\newcommand{\bsZ}{\hat{\mathbf{Z}}}
\newcommand{\bse}{\mbox{\bf{1}}}
\newcommand{\bspsi}{\hat{\boldsymbol{\psi}}}
\newcommand{\cdottt}{\!\cdot\!}
\newcommand{\deltaR}{\delta\mspace{-1.5mu}R}
\newcommand{\invup}{\rule{0ex}{2ex}}

\newcommand{\bGamma}{\boldmath$\Gamma$\unboldmath}
\newcommand{\dd}{\mbox{d}}
\newcommand{\ee}{\mbox{e}}
\newcommand{\p}{\partial}
\newcommand{\expmVo}{\langle\ee^{-{\mathbb V}}\rangle_0}

\newcommand{\Rav}{R_{\rm av}}
\newcommand{\Rc}{R_{\rm c}}

\newcommand{\la}{\langle}
\newcommand{\ra}{\rangle}
\newcommand{\rao}{\rangle\raisebox{-.5ex}{$\!{}_0$}}  
\newcommand{\rae}{\rangle\raisebox{-.5ex}{$\!{}_1$}}

\newcommand{\beq}{\begin{equation}}
\newcommand{\eeq}{\end{equation}}
\newcommand{\bea}{\begin{eqnarray}}
\newcommand{\eea}{\end{eqnarray}}
\def\lsim{\:\raisebox{-0.5ex}{$\stackrel{\textstyle<}{\sim}$}\:}
\def\gsim{\:\raisebox{-0.5ex}{$\stackrel{\textstyle>}{\sim}$}\:}

\numberwithin{equation}{section}

\thispagestyle{empty}
\title{\Large 
{\bf  Asymptotic statistics of the $n$-sided}\\[2mm]
{\bf planar Poisson-Voronoi cell: II. Heuristics}\\[3mm]
\phantom{xxx} }
 
\author{{H.\,J. Hilhorst}\\[5mm]
{\small Laboratoire de Physique Th\'eorique, B\^atiment 210}\\[-1mm] 
{\small Universit\'e Paris-Sud XI and CNRS}\\[-1mm]
{\small 91405 Orsay Cedex, France}\\}

\maketitle
\begin{small}
\begin{abstract}
\noindent
We develop a set of heuristic arguments to explain
several results on planar Poisson-Voronoi tessellations
that were derived earlier at the cost of considerable mathematical effort.
The results concern Voronoi cells having a large number $n$ of sides.
The arguments start from an entropy balance 
applied to the arrangement of $n$ neighbors around a central cell. 
It is followed by a simplified
evaluation of the phase space integral for the probability $p_n$
that an arbitrary cell be $n$-sided.
The limitations of the arguments are indicated.
As a new application we calculate the expected number of Gabriel
(or full) neighbors of an $n$-sided cell in the large-$n$ limit.\\

\end{abstract}
\end{small}
\vspace{45mm}

\noindent LPT Orsay 09-10\\
\thispagestyle{empty}
\newpage

%%%%%%%%%%%%%%%%%%%%%%%%%%%%%%%%%%%%%%%%%%%%%%%%%%%%%%%%%%%%%%%%%%%%%%%%%%%%%
%%%%%%%%%%%%%%%%%%%%%%%%%%%%%%%%%%%%%%%%%%%%%%%%%%%%%%%%%%%%%%%%%%%%%%%%%%%%%
%%%%%%%%%%%%%%%%%%%%%%%%%%%%%%%%%%%%%%%%%%%%%%%%%%%%%%%%%%%%%%%%%%%%%%%%%%%%%

\section{Introduction} 
\label{secintroduction}

Planar cellular structures occur in a wide variety of natural systems.
The examples most quoted are certain biological tissues 
and soap froths confined between parallel plates. 
In addition, cellular structures are employed 
as a tool of analysis in a great diversity of
problems throughout the sciences and beyond.
Many references may be found {\it e.g.} in Okabe {\it et al.} 
\cite{Okabeetal00} and in Rivier \cite{Rivier93}. 

\begin{figure}
\begin{center}
\scalebox{.45}
{\includegraphics{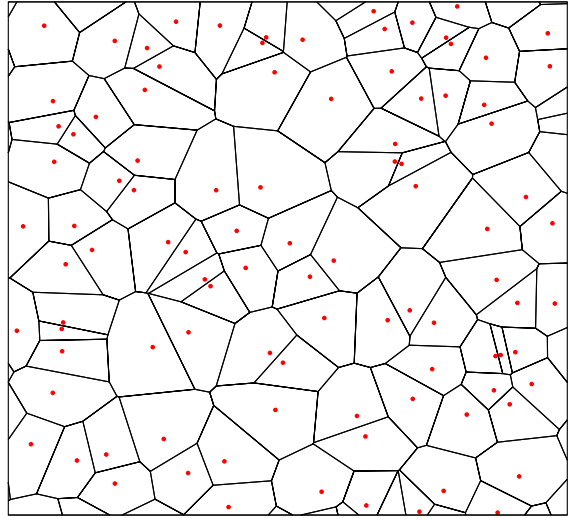}}
\end{center}
\caption{\small Voronoi tessellation of a set of point particles.
Voronoi cells are convex and their edges join at trivalent vertices.} 
\label{figVoronoi}
\end{figure}

The {\it Voronoi tessellation\,} is
one of the simplest mathematical models of a cellular
structure. 
In two dimensions it is obtained by distributing
a set of point-like ``seeds'' in the plane 
and then performing the Voronoi construction
as in the example of Fig.\,\ref{figVoronoi}: 
the plane is partitioned into cells such that each point of the
plane is in the cell of the seed to which it is closest.
Voronoi cells are convex and their edges join at trivalent vertices.
In the special case that the seed positions are drawn randomly from a uniform
distribution, one speaks of a {\it Poisson-}Voronoi tessellation.

%%%%%%%%%%%%%%%%%%%%%%%%%%%%%%%%%%%%
%%%%%%%%%%%%%%%%%%%%%%%%%%%%%%%%%%%%
%%%%%%%%%%%%%%%%%%%%%%%%%%%%%%%%%%%%
%%%%%%%%%%%%%%%%%%%%%%%%%%%%%%%%%%%%

The statistical
properties of the Poisson-Voronoi tessellation were studied
by Meijering \cite{Meijering53} as early as 1953.
Among the quantities of greatest interest
is the {\it sidedness probability\,} $p_n$, defined as the 
fraction of cells that are $n$-sided ($n=3,4,5,\ldots$).
As $n$ increases, $p_n$ passes through a maximum at $n=6$
and falls off to zero very rapidly for $n\gsim 10$.
The statistical properties of Voronoi cells range from fairly easy to very 
hard to determine, depending on the quantity of interest.
Many known properties are discussed and/or listed 
in Ref.\,\cite{Okabeetal00} (see, {\it e.g.}, Ch.\,5, Table 5.5.1).

It so happens that the sidedness probability  
$p_n$ is very hard to calculate.
Although it may readily be represented as a $2n$-dimensional integral,
the variables of integration
are coupled in such a way that this is a true many-particle
problem. As a result, no 
simple analytic result for $p_n$ is known for any $n$. 
We showed earlier, however,
that progress can be made if one considers the limit of large $n$.
In that limit $p_n$ is given by \cite{Hilhorst05a,Hilhorst05b}
\beq
{p}_n = \frac{C}{4\pi^2}\,\frac{(8\pi^2)^n}{(2n)!} 
\,\big[1+o(1)\big], \qquad n\to\infty,
\label{resultpn}
\eeq
where $ C = 
%\prod_{q=1}^\infty\,\left( 1-q^{-2}+ q^{-4} \right)^{-1}
0.344\,347...$\, \cite{footnoteb}.
Alternatively Eq.\,(\ref{resultpn}) may be written 
\beq
\log p_n = -2n\log n + n\log (2\pi^2\ee^2) -\tfrac{1}{2}\log n
- \tfrac{1}{2}\log \big( 2^6\pi^5 C^{-2} \big) 
+ {o}(1).
\label{resultlogpn}
\eeq
The interest of this asymptotic result, obtained by lengthy 
and technical mathematics,
goes much beyond the formula itself. Its derivation 
has brought along a full analysis
of the configurational statistics of the $n$-sided cell,
which in turn has opened the door to further developments in various 
directions that we will briefly mention now. 
\vspace{2mm}

(i) First, let $m_n$ be the average sidedness of the {\it neighbor\,} of an
$n$-sided cell.
In the wake of the work of Ref.\,\cite{Hilhorst05b}
it was shown \cite{Hilhorst06} that  
this sidedness pair correlation 
has the large-$n$ expansion $m_n=4+3{\pi^\half}{n^{-\half}}+\ldots$.
This demonstrated that Aboav's celebrated law \cite{Aboav70}, 
usually written as $nm_n=an+b$, is in fact a linear approximation useful for, 
but limited to, small $n$ values.

(ii) Secondly, 
the development of a new Monte Carlo method \cite{Hilhorst07}
made it possible to simulate $n$-sided cells 
and to numerically determine $p_n$
with four-digit precision for all finite values of $n$.
This method was used to generate snapshots
of extremely large cells with sidednesses as high as $n=1600$ \cite{footnotee}.

(iii) Thirdly, the analytic methods developed for Voronoi cells in
Refs.\,\cite{Hilhorst05a,Hilhorst05b} proved to be of wider use.
In Ref.\,\cite{HilhorstCalka08} 
they were applied to line tessellations of the plane and linked up with work 
in mathematics by Hug and Schneider \cite{HugSchneider07}.
In Ref.\,\cite{Hilhorstetal08} they were brought to bear
on {\it Sylvester's question\,} \cite{Sylvester1864}: 
what is the probability
$p^*_n$ that $n$ points chosen randomly from a uniform distribution in a disk,
are the vertices of an $n$-sided convex polygon? 
This question leads to a connection between on the one hand
our studies of many-sided Voronoi cells, and on the other hand
work on random points in convex position by B\'ar\'any
\cite{Barany99}, B\'ar\'any {\it et al.} \cite{Baranyetal00}, and Calka and
Schreiber \cite{CalkaSchreiber05}, 
as well as work in statistical physics
on extremal statistics of random walks 
by Gy\"orgyi {\it et al.} \cite{Gyorgyietal07} and Majumdar and Comtet 
\cite{MajumdarComtet04,MajumdarComtet05}.
\vspace{2mm}

Because of all these ramifications we are justified to ask 
which of the results on many-sided Voronoi cells 
can be understood without the full
mathematical apparatus that was necessary in Ref.\,\cite{Hilhorst05b}
and in most of the following articles.
This paper answers that question.
It is a successor to Ref.\,\cite{Hilhorst05b}, 
of which we will reproduce part of the results by new, elementary 
but heuristic arguments. We also indicate the limitations of these arguments.
Finally, the present paper will lay the basis 
for a study of Voronoi cells in higher spatial dimensions \cite{Hilhorst09}. 
\vspace{2mm}

Our presentation is as follows.
A preliminary observation in section \ref{secobservation} has to do with the
difference between the statistics of {\it large\,} and of {\it many-sided\,}
cells. We then estimate $p_n$ in two successive stages.
First, in section \ref{secestimatepn}, and considering the many-sided cell as
circular, we present heuristic arguments
leading to the simplest of all possible estimates.
Secondly, in section \ref{secimproved}, and building on what we have learned, 
we perform part of the mathematics of Ref.\,\cite{Hilhorst05b} 
in a strongly simplified way.
This leads to a much improved estimate of $p_n$.

In section \ref{secelastic} we consider 
an effect not captured by the preceding arguments,
namely the ``elastic'' deformations of the cell perimeter from circularity.
The key to their analysis is the so-called
``random acceleration process,'' also known from the theory of Brownian motion.
We derive the scaling with $n$ of the elastic deformations;
it is not possible, however, to obtain heuristically
their precise contribution to $p_n$.

In section \ref{secGabriel}, finally, we consider the average number 
of Gabriel neighbors of an $n$-sided cell 
and show that for large $n$ it scales as $\sim n^\half$. 
Adjacent Voronoi cells are called Gabriel 
neighbors when the line segment
that connects their seeds does not intersect a third cell.
Gabriel neighbors have been studied in mathematics
\cite{Moller94,MollerStoyan07} and also play an important role in certain
pattern recognition algorithms where ``decision boundaries'' must be
constructed \cite{Bhattacharyaetal05,ChenWei06}.
Section \ref{secconclusion} is our conclusion.

%%%%%%%%%%%%%%%%%%%%%%%%%%%%%%%%%%%%%%%%%%%%%%%%%%%%%%%%%%%%%%%%%%%%%%%%%%%%%
%%%%%%%%%%%%%%%%%%%%%%%%%%%%%%%%%%%%%%%%%%%%%%%%%%%%%%%%%%%%%%%%%%%%%%%%%%%%%

\section{A preliminary observation}
\label{secobservation}

We consider a Poisson-Voronoi tessellation with seed density $\rho$; 
a measure for the typical interparticle distance is therefore  
$\ell_{\rm ip}=\rho^{-\half}$.
For all questions of interest this density may be scaled to unity, 
but we will keep it as a check on dimensionality.
We select an arbitrary seed, let its position be the origin, and 
are interested in its Voronoi cell, which we will call the ``central cell.''
This cell has the same statistical properties as all others.
Our focus is, first of all, on its sidedness probability $p_n$. 

%%%%%%%%%%%%%%%%%%%%%%%%%%%%%%%%%%%%%%%%%%%%%%%%%%%%%%%%%%%%%%%%%%%%%%%%%%%%%

\subsection{Large cells}
\label{seclargecells}

It is useful to begin by pointing out a distinction. 
When the central 
cell is {\it large,} in the sense of having a large area, it will also
typically have many sides, and inversely.
Nevertheless, the statistical
subensemble of central cells with a prescribed area $A$
is inequivalent to the one where it has a prescribed number $n$ of sides,
even in the limit of large $A$ and $n$. We will now discuss why.

%%%%%%%%%%%%%%%%%%%%%%%%%%%%%%%%%%%%
%%%%%%%%%%%%%%%%%%%%%%%%%%%%%%%%%%%%
\begin{figure}
\begin{center}
\scalebox{.45}
{\includegraphics{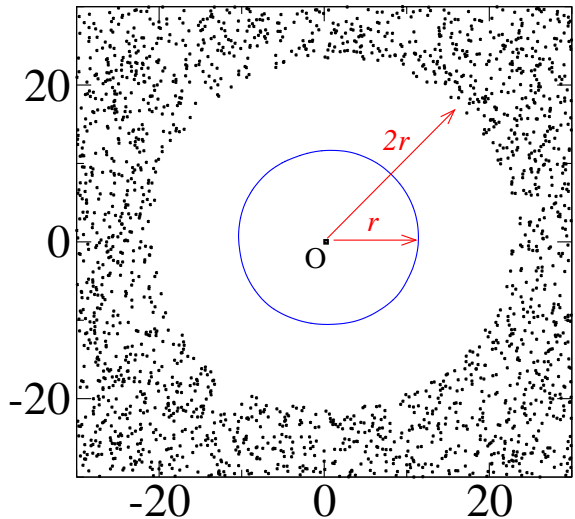}}
\end{center}
\caption{\small Black dots: seeds randomly distributed in the plane with an
  {\it a priori\,} uniform density $\rho=1$. The picture shows a
  very low-probability configuration consisting of a seed in the origin
  surrounded by a disk of radius $2r$ void of other seeds.
  The Voronoi cell of the central seed will be a convex 
  $n_r$-sided polygon, where $n_r$ is a random number 
  of order $r^{\frac{2}{3}}$.
  The cell perimeter will roughly coincide with the circle of radius $r$ shown
  in the figure.}
\label{figemptydisk}
\end{figure}
%%%%%%%%%%%%%%%%%%%%%%%%%%%%%%%%%%%%
%%%%%%%%%%%%%%%%%%%%%%%%%%%%%%%%%%%%

Let us first ask how it can happen that the central cell
(say an approximately circular one \cite{footnotea}) has a large 
area $A=\pi r^2$.
As shown in Fig.\,\ref{figemptydisk}, this occurs when
a disk of radius $2{r}$ 
around the central seed is free of other seeds.
Now let us inquire about the sidedness $n_{r}$ of this large cell, or,
equivalently, its number of first-neighbor cells.
One might think naively that in the limit of large $r$ this number
can be estimated as the product of the seed density $\rho$ and the area of
an annulus of width $\ell_{\rm ip}$ along the perimeter of the empty disk.
That would give $n_{{r}}\sim \rho\times\ell_{\rm ip}\times 4\pi{r}
\sim \rho^\half{r}$, 
where the sign \,$\sim$\, denotes asymptotic proportionality as $r$
gets large.
The Voronoi construction, however, does not confirm this
linearity between $n_r$ and $r$.
Thanks to work by Calka \cite{Calka02} and Calka and Schreiber 
\cite{CalkaSchreiber05} we know that in fact the 
number of first neighbors increases only as
$n_{r}\sim(\rho^{\half}{r})^{\frac{2}{3}}$ and that they
are essentially located in a circular annulus 
of width $\sim \rho^{-\frac{2}{3}}{r}^{-\frac{1}{3}}$ 
along the perimeter of the empty disk.
The cause is a screening effect that we will study in detail in
section \ref{secscreening}. 

%%%%%%%%%%%%%%%%%%%%%%%%%%%%%%%%%%%%%%%%%%%%%%%%%%%%%%%%%%%%%%%%%%%%%%%%%%%%%

\subsection{Constructing a many-sided cell}
\label{secfromlarge}

Obtaining an $n$-sided cell by the construction of Fig.\,\ref{figemptydisk}
would require having an empty disk of radius $2r$
such that $n_{r}=n$, and hence would require 
$r \sim {r}_n\equiv\rho^{-\half}n^{\frac{3}{2}}$.
The probability for this to happen is 
$\exp(-\pi\rho{r}_n^2)=\exp(-\mbox{cst}\times n^3)$,
which with respect to an arbitrary seed distribution
corresponds to an entropy loss $\Delta S$ equal to \cite{footnoteg}
\beq
\Delta S \sim -n^3, \qquad n\to\infty.
\label{exptentrdefic}
\eeq
It then becomes clear that perhaps there is a way of creating an $n$-sided
central cell at lower entropy cost. 
If the annulus with the $n$ first neighbors is contracted homothetically, 
the central cell remains $n$-sided. 
The reduction of the annular surface costs entropy
but there is a compensation due to
the extra space that becomes available to the seeds 
outside the annulus. 
In the next section we will obtain the 
maximum-entropy arrangement by balancing these two effects against each other
in what may be called an {\it entropy-versus-entropy\,} argument. 

%%%%%%%%%%%%%%%%%%%%%%%%%%%%%%%%%%%%%%%%%%%%%%%%%%%%%%%%%%%%%%%%%%%%%%%%%%%%%
%%%%%%%%%%%%%%%%%%%%%%%%%%%%%%%%%%%%%%%%%%%%%%%%%%%%%%%%%%%%%%%%%%%%%%%%%%%%%

\section{Heuristic estimate for $p_n$}  
\label{secestimatepn}

In order to heuristically estimate the sidedness probability $p_n$ we proceed
in two stages: a description of the screening effect in section
\ref{secscreening} and an entropy balance in section \ref{secestimating}.

%%%%%%%%%%%%%%%%%%%%%%%%%%%%%%%%%%%%%%%%%%%%%%%%%%%%%%%%%%%%%%%%%%%%%%%%%%%%%

\subsection{Screening}
\label{secscreening}

%%%%%%%%%%%%%%%%%%%%%%%%%%%%%%%%%%%%
%%%%%%%%%%%%%%%%%%%%%%%%%%%%%%%%%%%%
\begin{figure}
\begin{center}
\scalebox{.45}
{\includegraphics{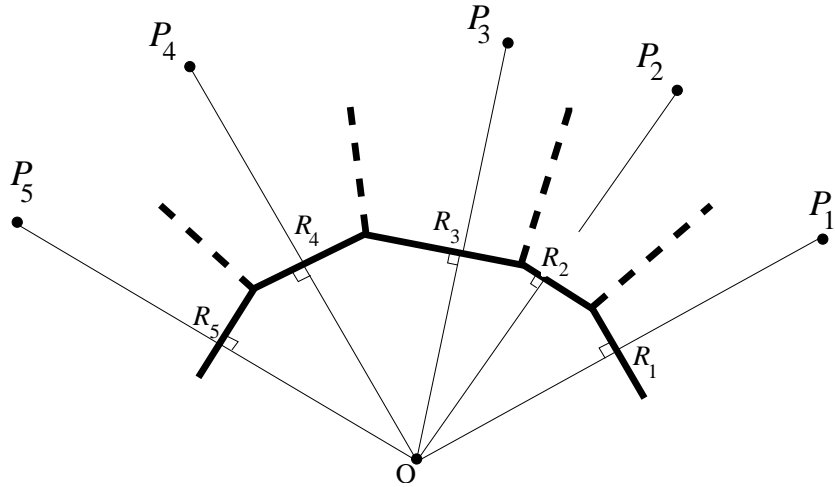}}
\end{center}
\caption{\small A central seed located in the origin $O$ and five of its
  neighboring seeds at positions $P_1,\ldots,P_5$. The ``mid-points''
  $R_1,\ldots,R_5$ are the centers of the intervals $OP_1,\ldots,OP_5$.
  Heavy solid line: part of the perimeter of the central Voronoi cell. Heavy
  dashed lines: boundaries between the first-neighbor Voronoi cells.} 
\label{figperimeter1}
\end{figure}
%%%%%%%%%%%%%%%%%%%%%%%%%%%%%%%%%%%%
%%%%%%%%%%%%%%%%%%%%%%%%%%%%%%%%%%%%

\begin{figure}
\begin{center}
\scalebox{.45}
{\includegraphics{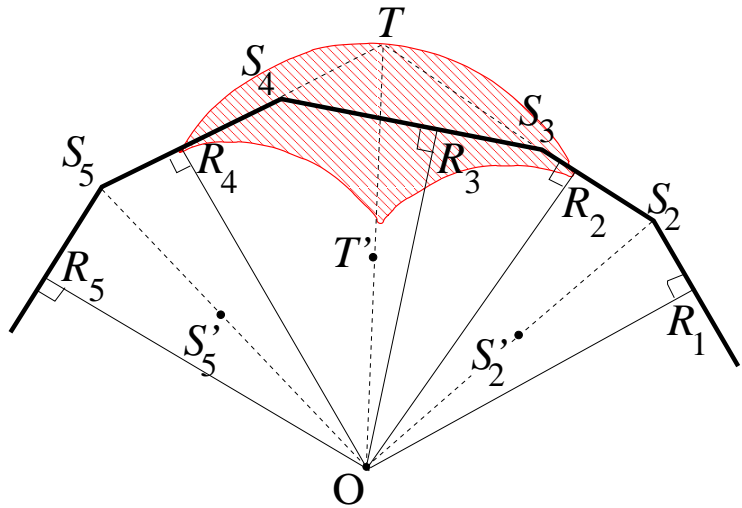}}
\end{center}
\caption{\small Heavy solid line:
same part of the perimeter of the central Voronoi cell as
shown in Fig.\,\ref{figperimeter1}.
The vertices are denoted $S_2,\ldots,S_5$; point
$T$ is a virtual vertex: it would be there if the 
perimeter segment $S_3S_4$ were absent.
The shaded region is the area available to mid-point $R_3$ 
when all other mid-points remain fixed and
the number of cell sides is not allowed to change. This region is
bounded by three circular arcs; the circles pass through the origin and
have their centers in $S'_2,\, T'$, and $S'_5$
halfway the intervals $OS_2$, $OT$, and $OS_5$, respectively.} 
\label{figperimeter2}
\end{figure}

Fig.\,{\ref{figperimeter1}} 
shows the central seed in $O$ together with five of its
first-neighbor seeds located at $P_1,\ldots,P_5$.
By constructing in the ``mid-points'' $R_1,\ldots,R_5$
the perpendicular bisectors of $OP_1,\dots,OP_5$ one obtains
five segments of the perimeter of the central cell.
The positions of the vertices $S_2,\ldots,S_5$
then follow as indicated in  Fig.\,\ref{figperimeter2}.
We will refer to the line interval $S_{m}S_{m+1}$ as the $m$th 
perimeter segment (with cyclic boundary conditions, $S_{n+1} \equiv S_1$).
Obviously, integrating over the $n$ nearest-neighbor positions
$P_1,\ldots,P_n$ is the same thing, apart from
trivial factors 2, as integrating over the $n$ mid-point positions
$R_1,\ldots,R_n$.

%%%%%%%%%%%%%%%%%%%%%%%%%%%%%%%%%%%%
%%%%%%%%%%%%%%%%%%%%%%%%%%%%%%%%%%%%
%%%%%%%%%%%%%%%%%%%%%%%%%%%%%%%%%%%%
%%%%%%%%%%%%%%%%%%%%%%%%%%%%%%%%%%%%

We ask now the following question: what is the area available to an 
arbitrary seed if all other seed positions as well as the number $n$ of
perimeter segments are kept fixed?
Specifically, let us study in Fig.\,\ref{figperimeter2} how $R_3$ can move
around when $R_1, R_2, R_4, R_5$ are fixed.
Both the polar angle and the length of the vector $OR_3$
may vary, but $S_3S_4$ should stay perpendicular to this vector.
The domain of variation of $R_3$ is, clearly,
determined by the condition that the
intersection points $S_3$ and $S_4$ remain confined to the 
immobile intervals $S_2T$ and $S_5T$, respectively.
Some reflection then shows that $R_3$ is restricted to the shaded region in
Fig.\,\ref{figperimeter2}. This region is bounded by three arcs of
circles centered at points $S'_2,\, T'$, and $S'_5$ which are located 
halfway between the origin and the corresponding vertices.
The exact surface area of the shaded region may be expressed in terms of the
mid-point coordinates, but we will not need that much detail.

%%%%%%%%%%%%%%%%%%%%%%%%%%%%%%%%%%%%
%%%%%%%%%%%%%%%%%%%%%%%%%%%%%%%%%%%%
\begin{figure}
\begin{center}
\scalebox{.45}
{\includegraphics{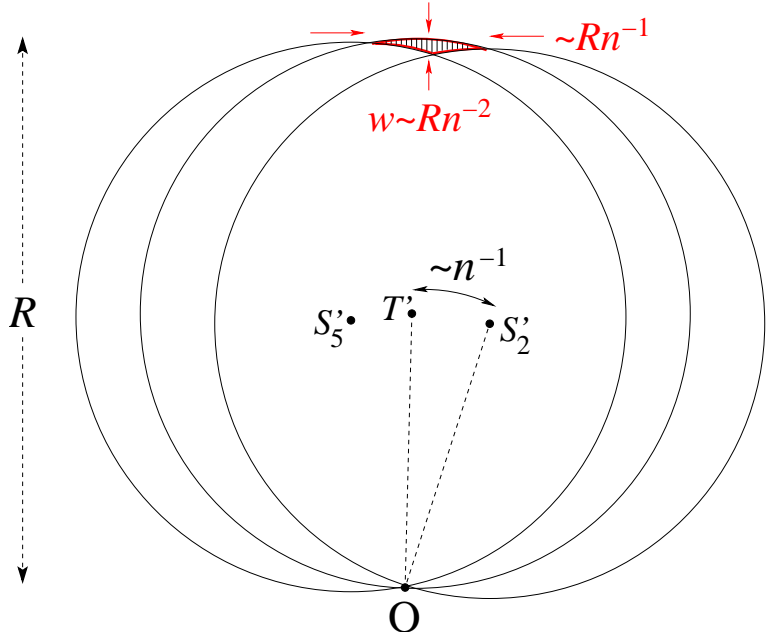}}
\end{center}
\caption{\small 
The three circles whose arcs were shown in Fig.\,\ref{figperimeter2}, 
are represented here in full and for much larger $n$.
Their diameters are approximately equal to $R$ and their centers
form angles or order $n^{-1}$ with the origin.
In the limit of large $n$ the shaded area 
becomes triangular with two acute angles of order $n^{-1}$.} 
\label{fig3circles}
\end{figure}
%%%%%%%%%%%%%%%%%%%%%%%%%%%%%%%%%%%%
%%%%%%%%%%%%%%%%%%%%%%%%%%%%%%%%%%%%

In the limit of large $n$, shown in Fig.\,\ref{fig3circles}, 
the angles $S_2^\prime OT^\prime$ and $S_5^\prime OT^\prime$ 
will typically be of order $n^{-1}$.
The three circles have been represented in full in this figure.
Let $R$ be their typical diameter (which is also the {\it radius\,}
of the central cell). One easily shows that for large $n$ 
the shaded area tends towards a triangle
with a long side of order $Rn^{-1}$ and two acute angles of order $n^{-1}$.
It follows that the triangle height $w$ scales as
\beq
w\sim Rn^{-2}, \qquad n\to\infty.
\label{heurw}
\eeq
This equation expresses the screening effect: the larger $n$, the narrower the
annulus containing the $n$ points. We will make essential use of 
relation (\ref{heurw}) in the subsection \ref{secestimating}. 
%%% PVC 3601 for estimate of intrinsic width
It is worth noting the following consequence. 
The shaded area in Fig.\,\ref{figperimeter2} shows that  
when the polar angle of $R_3$ approaches the one of $R_{2}$ or $R_{4}$,
the available phase space goes to zero linearly in the angle difference.  
This means that the screening leads to an {\it effective repulsion between the
mid-points,} an effect that 
will be made more quantitative in section \ref{secimproved}. 

%%%%%%%%%%%%%%%%%%%%%%%%%%%%%%%%%%%%%%%%%%%%%%%%%%%%%%%%%%%%%%%%%%%%%%%%%%%%%

\subsection{Entropy balance}
\label{secestimating}

%%%%%%%%%%%%%%%%%%%%%%%%%%%%%%%%%%%%
%%%%%%%%%%%%%%%%%%%%%%%%%%%%%%%%%%%%
\begin{figure}
\begin{center}
\scalebox{.45}
{\includegraphics{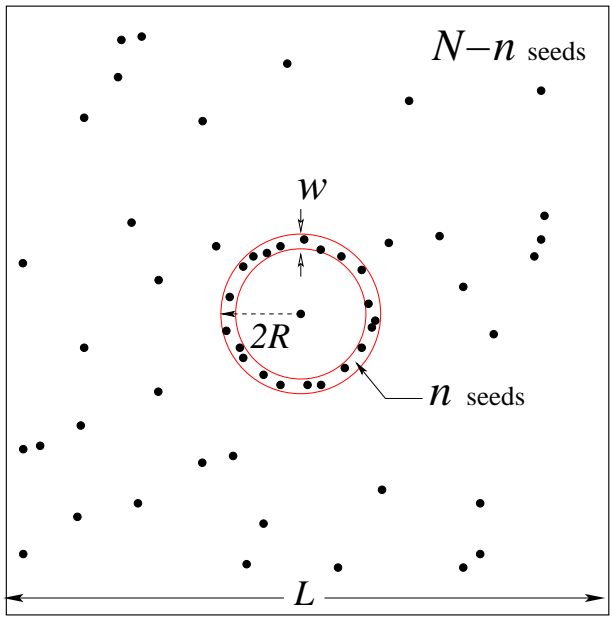}}
\end{center}
\caption{\small Given an annulus of outer radius $2R$ and width $w$
surrounding a central seed, we
ask for the probability 
that out of $N$ other seeds in an area $L^2$,
there be none in the inner disk of radius $2R-w$ and
exactly $n$ inside the annulus, the remaining ones occupying the outer
region.} 
\label{figenviron}
\end{figure}
%%%%%%%%%%%%%%%%%%%%%%%%%%%%%%%%%%%%
%%%%%%%%%%%%%%%%%%%%%%%%%%%%%%%%%%%%

\begin{figure}
\begin{center}
\scalebox{.45}
{\includegraphics{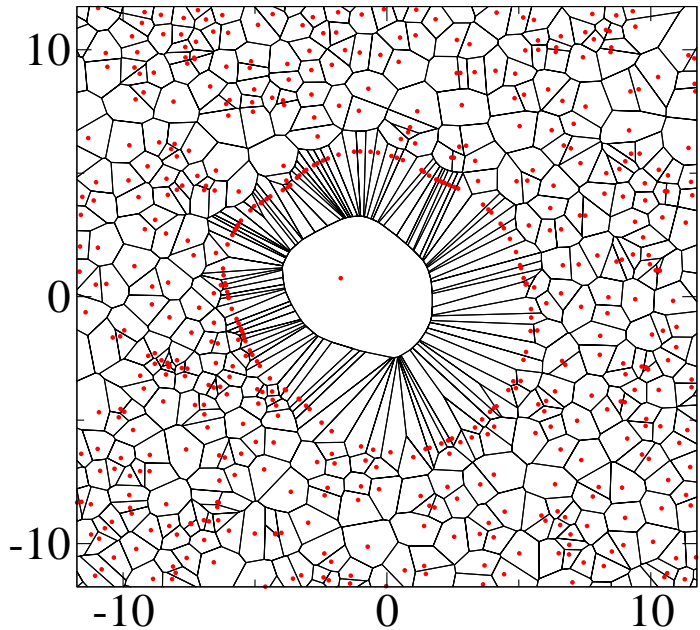}}
\end{center}
\caption{\small A Poisson-Voronoi tessellation containing a cell
of $n=96$ sides.
The seed density is $\rho=1$. 
One clearly distinguishes the alignment of the first-neighbor seeds 
along an almost continuous curve. The deviations of this curve from circularity
(see section \ref{secelastic}), predicted
to disappear when $n\to\infty$, are still important 
for this value of $n$.} 
\label{figcell96}
\end{figure}

The considerations above suggest the next question.
We consider a central seed in the origin and $N$ others 
in an area $L\times L$. At a suitable point we will let
$N,L\to\infty$ with $\rho\equiv N/L^2$ fixed.
We now ask what the probability 
is for the seed configuration to be constrained as
illustrated in Fig.\,\ref{figenviron}: of all seeds,
$n$ are located in an annulus of outer radius $2R$ and width $w$, and the
remaining ones are all outside of this annulus.
This probability will be our approximation for the sidedness
probability $p_n$ defined above and we will denote it by the same symbol $p_n$.
We may write it as the exponential of an entropy difference,
\beq
{p}_n=\ee^{S'-S}=\ee^{\Delta S},
\label{defpbarn}
\eeq
where ${S'}$ and $S$ are the entropies of the constrained and
unconstrained seed configuration, respectively. 
Let $a_1$ and $a_2$ denote the area of the annulus and the area of the
inner disk plus annulus, respectively. As functions of the two lengths
$R$ and $w$ they read
\bea
a_1&=&\pi(2R)^2-\pi(2R-w)^2, \nonumber\\[2mm]
a_2&=&\pi(2R)^2.
\label{defa1a2}
\eea
In terms of these two areas we have
\beq
\ee^{S}=V^N, \qquad \ee^{S'}=\binom{N}{n}\left( L^2-a_2 \right)^{N-n} a_1^n.
\label{exprSSp}
\eeq
Using (\ref{exprSSp}) in (\ref{defpbarn}) and taking the limit
$N,L\to\infty$ as stated we get
\beq
{p}_n=\frac{(\rho a_1)^n}{n!}\,\ee^{-\rho a_2}.
\label{expr1pbarn}
\eeq
Taking the logarithm of (\ref{expr1pbarn}) and using (\ref{defa1a2})
we obtain
\bea
\log {p}_n &=& -\log n! + n\log\rho a_1 - \rho a_2 \nonumber\\[2mm]
&=& -\log n! +n\log\,[\rho\pi w(4R-w)] - \rho\pi(2R)^2.
\label{expr1logpbar}
\eea
At this point we will take into account the screening effect. 
Motivated by the discussion of section \ref{secscreening}
we impose that $w=2cRn^{-2}$,
where $c$ is a numerical constant. This ensures in some average sense 
that the seeds in the annulus are not screening each other.
Eliminating $w$ from (\ref{expr1logpbar})
in favor of $R$ and $n$ we find
\beq
\log{p}_n = -\log n! +n\log (8\pi\rho cR^2n^{-2}) - 4\pi\rho R^2
%+{\cal O}(R^2n^{-2})
+{\cal O}(n^{-1}), 
\label{expr2logpbar}
\eeq
valid in the limit $n\to\infty$. 
Eq.\,(\ref{expr2logpbar}) represents the entropy loss
associated with the constraints on the seed positions.
The only freedom that we have left is to choose the relation 
between $R$ and $n$.
Varying (\ref{expr2logpbar}) with respect to $R$
shows that it has a minimum for $R=R^*$, where
\beq
R^*=\left( \frac{n}{4\pi\rho} \right)^{\!\frac{1}{2}}.
\label{exprRstar}
\eeq
This expression for $R^*$, 
obtained here heuristically, coincides with the exact result of
Refs.\,\cite{Hilhorst05a,Hilhorst05b} for the radius $R_{\rm c}$  
of the $n$-sided Voronoi cell.
The annular width $w^*$ corresponding to the maximum entropy
is obtained by substitution of (\ref{exprRstar}) in (\ref{heurw}),
\beq
w^*=c(\pi\rho)^{-\half}n^{-\frac{3}{2}}.
\label{exprwstar}
\eeq
Eqs.\,(\ref{exprRstar}) and (\ref{exprwstar})
show that as $n$ grows, the radius $2R^*$
of the annulus of first-neighbor seeds increases 
but its width $w^*$ decreases. 
The typical distance $\ell^*$
between two adjacent first neighbors along this circle
is $\ell^*=4\pi R^*/n=(4\pi/\rho n)^\half \sim n^{-\half}$.
From the fact that $w^*\sim n^{-\frac{3}{2}}$ and hence $w^* \ll \ell^* \ll 1$
it follows that in the large $n$ limit the first order neighbors align 
along an almost continuous curve.
This is confirmed by independent Monte Carlo simulation according to the
method of Ref.\,\cite{Hilhorst07}. Fig.\,\ref{figcell96} shows a
Poisson-Voronoi tessellation containing a $96$-sided cell. 
The alignment of the first-neighbor seeds is clearly visible. 
This figure shows still another feature of such a large Voronoi cell:
although the cell itself is convex, this need not be the case for
curve of alignment of the first-neighbor seed positions.

This is also the suitable place for another side comment.
When the ``seeds'' are physical particles having some nonzero diameter
$a$, obviously the present considerations cease to be valid when
the density of particles becomes too high. From the preceding discussion 
it is clear that an appropriate criterion for their validity
is that $a \ll \ell^*$. 
This point was briefly discussed in Ref.\,\cite{Hilhorst08}.
\vspace{2mm}

%%%%%%%%%%%%%%%%%%%%%%%%%%%%%%%%%%%%
%%%%%%%%%%%%%%%%%%%%%%%%%%%%%%%%%%%%
%%%%%%%%%%%%%%%%%%%%%%%%%%%%%%%%%%%%
%%%%%%%%%%%%%%%%%%%%%%%%%%%%%%%%%%%%

To complete the calculation of this section
we determine the expression for the sidedness probability
as given by the present approximation.
Substitution of (\ref{exprRstar}) in (\ref{expr2logpbar})
gives
\bea
\log{p}_n &=& -\log n! +n\log 2cn^{-1} -n +\ldots \nonumber\\[2mm]
&=& -2n\log n +n\log 2c -\thalf\log n -\thalf\log 2\pi + \ldots, 
\label{exprlogbarpstar}
\eea
in which the dot terms vanish as $n\to\infty$. 

We now compare
Eq.\,(\ref{exprlogbarpstar}) to the exact expansion (\ref{resultlogpn}). 
The leading order term of (\ref{exprlogbarpstar}) is correct and
this is the first time that a heuristic argument achieves this.
We recall that Drouffe and Itzykson
\cite{DrouffeItzykson84,ItzyksonDrouffe89} derived the lower bound 
$p_n=-\alpha n\log n+\ldots$ with
$\alpha\leq 2$; their conjecture was that $\alpha=2$, which we now know 
\cite{Hilhorst05a,Hilhorst05b} to be correct \cite{footnoteb}.
The second term in expansion (\ref{exprlogbarpstar}) differs from the 
exact result (\ref{resultlogpn}).
Indeed, obtaining this second term correctly requires a more
detailed study of the arrangement of the $n$ seeds in the annulus.
This will be the subject of section \ref{secimproved}.

%%%%%%%%%%%%%%%%%%%%%%%%%%%%%%%%%%%%%%%%%%%%%%%%%%%%%%%%%%%%%%%%%%%%%%%%%%%%%
%%%%%%%%%%%%%%%%%%%%%%%%%%%%%%%%%%%%%%%%%%%%%%%%%%%%%%%%%%%%%%%%%%%%%%%%%%%%%

\section{Improved estimate for $p_n$}
\label{secimproved}

Building on the experience gained above
we can now greatly improve upon the calculation of $p_n$.
In this section $\bP_m$ will denote the position 
vector of the $m$th first-neighbor seed and $P_m\equiv |\bP_m|$
will stand for its length, with $m=1,2,\ldots,n$.
All exact work on $p_n$ should start from 
the $2n$-fold phase space integral
\beq
p_n=\frac{\rho^n}{n!}\int\!\dd\bP_1\ldots\dd\bP_n\,\chi\,\ee^{-\rho{\cal A}},
\label{basicpn}
\eeq
where $\chi=1$ (or $\chi=0$) if the condition
for the $n$ seed positions $\bP_m$ to each contribute a perimeter segment
is satisfied (or not satisfied); 
and where ${\cal A}$ is the area from which the remaining 
seeds must be excluded in order not to interfere with this $n$-sided cell.
In this section we will evaluate (\ref{basicpn})
in an approximation suggested by the considerations 
of section \ref{secestimatepn}. 
The expression $\chi\exp(-\rho{\cal A})$
is a complicated function of the variables of integration
that we will not need in detail in the present approach. 

%%%%%%%%%%%%%%%%%%%%%%%%%%%%%%%%%%%%%%%%%%%%%%%%%%%%%%%%%%%%%%%%%%%%%%%%%%%%%%

\subsection{The $2n$-fold integral for $p_n$}
\label{secintegration}

We assume that for large $n$
the shape of the $n$-sided Voronoi cell will be close to a circle.
Let a new variable $P$ denote twice this circle's radius
and hence be representative of the linear scale of the cell.
We can then split the $2n$-fold integration in (\ref{basicpn})
into a single one on $P$ and a set of $2n-1$ integrations that define
the detailed shape and the orientation of the cell. For this set
we will take the $n$ polar angles of the $\bP_m$ and the $n-1$
independent ratios $P_{m}/P_{m-1}$,\,\, $m=2,\ldots,n$ 
of their radial distances. These ratios may in turn be expressed in terms of
angular variables. 
Assuming that with negligible error 
we may factorize the integral on $P$, to be 
denoted $I_{\rm rad}$, and the angular
ones, we then have
\beq
p_n=\frac{1}{n!}\,I_{\rm rad}I_{\rm ang}\,,
\label{splitpn}
\eeq
in which we will consider $I_{\rm rad}$ and $I_{\rm ang}$ separately.

All seeds other than the $n$ first neighbors
must stay outside the disk of radius $P$, 
which happens with probability $\exp(-\rho\pi P^2)$. 
Hence, extracting a scale factor $P$ from all
$2n$ position variables in the integration (\ref{basicpn}), we have
\beq
I_{\rm rad} = \rho^n\!\int_0^\infty\! \dd P\, P^{2n-1}\,\ee^{-\rho\pi P^2}
= \frac{(n-1)!}{2\pi^n}\,.
\label{radialint}
\eeq
When this integral is performed, the seed density $\rho$ disappears 
from the expression for $p_n$, as it had to.

\begin{figure}
\begin{center}
\scalebox{.45}
{\includegraphics{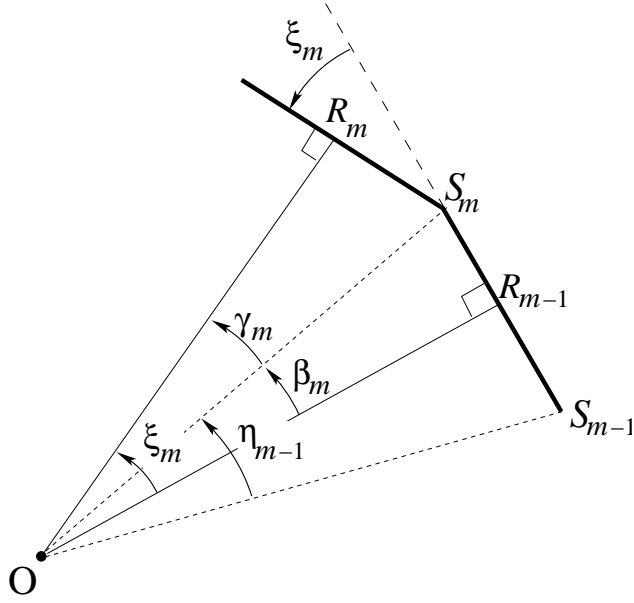}}
\end{center}
\caption{\small Definition ($n$-cyclic in the index $m$) 
of the angles of integration $\xi_m$ and $\eta_m$
and of the auxiliary angles $\beta_m$ and $\gamma_m$.} 
\label{figtransformation}
\end{figure}

The integral $I_{\rm ang}$ in (\ref{splitpn}) requires more discussion.
It involves $2n-1$ dimensionless variables of integration
for which we can take any set of $2n-1$ independent angles.
Introducing an extra factor $(n-1)!$ allows us to order 
the $n$ seeds such that their indices correspond to increasing polar angles.
A natural choice of angles is shown in Fig.\,\ref{figtransformation}: 
it is the set of polar angle differences $\{\xi_m,\eta_m\}$,
where $\xi_m$ is the angle between two successive mid-point vectors 
$R_{m-1}$ and $R_m$\,, and
$\eta_m$ the angle between two successive vertex vectors $S_m$ and $S_{m+1}$.

Fig.\,\ref{figtransformation} also illustrates how
the perimeter segments may be constructed successively
when the $\xi_m$ and $\eta_m$ are given.
Two types of construction steps alternate.
First, given $S_{m-1}$ and a direction of departure,
continue along this direction
until the segment $S_{m-1}S_m$ spans the angle $\eta_{m-1}$; 
secondly, when arriving at $S_m$, define a new direction of departure 
by taking an angle of $\xi_m$ to the left. 
Iterate $n$ times.

%%%%%%%%%%%%%%%%%%%%%%%%%%%%%%%%%%%%
%%%%%%%%%%%%%%%%%%%%%%%%%%%%%%%%%%%%
%%%%%%%%%%%%%%%%%%%%%%%%%%%%%%%%%%%%
%%%%%%%%%%%%%%%%%%%%%%%%%%%%%%%%%%%%

It is easy to introduce the $\xi_m$ into (\ref{basicpn}) 
by passing to a polar coordinate
representation for the $\bP_m$;
however, the passage from the ratios $P_{m}/P_{m-1}$ to the $\eta_m$ 
leads to the appearance of a nontrivial
Jacobian that we will consider now.
It follows from Fig.\,\ref{figperimeter1} and Fig.\,\ref{figtransformation},
where we have introduced an auxiliary angle $\beta_m$, that
\beq
\frac{P_m}{P_{m-1}} = \frac{|OR_m|}{|OR_{m-1}|} = 
\frac{\cos(\xi_m-\beta_m)}{\cos\beta_m}
= \cos\xi_m-\sin\xi_m\tan\beta_m\,.
\label{transf}
\eeq
Fig.\,\ref{figtransformation} shows that the angles $\eta_{m-1}$ and
$\beta_m$ vary with the position of $S_m$ at a common rate. 
To find the Jacobian we therefore calculate the derivative 
\beq
\frac{\dd (P_m/P_{m-1})}{\dd\eta_{m-1}} 
\,=\, \frac{\sin\xi_m}{\cos^2\beta_m} \,\simeq\, \xi_m\,,
\label{exprjac}
\eeq
the last equality being valid asymptotically for small
$\xi_m$ and $\beta_m$, which is the relevant limit when $n$ gets large.
By going around the perimeter one thus finds that the transformation from the
$P_{m}/P_{m-1}$ to the $\eta_m$ is accompanied by a Jacobian\,
$\xi_1\xi_2\ldots\xi_n$. 
This result quantifies our
observation at the end of section \ref{secscreening},
namely that there is an effective repulsion between the mid-points.

The integral on the angular variables now becomes 
%%% [*** This is where we forget about the no-spiral condition.] 
\beq
I_{\rm ang}=(n-1)!\,I_{\rm vert}I_{\rm midp}\,,
\label{splitIang}
\eeq
in which 
\beq
I_{\rm vert}=\int_0^{2\pi}\! \dd\eta_1\ldots\dd\eta_n\,
       \delta\left(\summ\eta_m-2\pi\right)
=\frac{(2\pi)^{n-1}}{(n-1)!}\,,
\label{exprIvert}
\eeq
\beq
I_{\rm midp}=\int_0^{2\pi}\! \dd\xi_1\ldots\dd\xi_n\,(\xi_1\ldots\xi_n)\,
       \delta\left(\summ\xi_m-2\pi\right)
=\frac{(2\pi)^{2n-1}}{(2n-1)!}\,,
\label{exprImidp}
\eeq
where the delta functions enforce obvious sum rules. 
Due to these constraints the angles $\xi_m$ and $\eta_m$ will typically
take values of order $n^{-1}$ with 
average $2\pi/n$.
The exact value (here $2\pi$) of the
upper integration limits in (\ref{exprIvert}) and (\ref{exprImidp})
does not affect the outcome of the integrals
to the order in $n$ that is relevant for our discussion.
%%% [*** PVC931]

%%%%%%%%%%%%%%%%%%%%%%%%%%%%%%%%%%%%%%%%%%%%%%%%%%%%%%%%%%%%%%%%%%%%%%%%%%%%%

\subsection{Improved result}
\label{secsubimproved}

We now know all the factors that go into the final result.
Substituting 
(\ref{exprIvert}) and 
(\ref{exprImidp}) in (\ref{splitIang}) and using the result 
together with (\ref{radialint}) in (\ref{splitpn}) gives
\bea
p_n&=&\frac{1}{n!}
\times (n-1)!
\times\frac{(n-1)!}{2\pi^n}
\times\frac{(2\pi)^{n-1}}{(n-1)!}
\times\frac{(2\pi)^{2n-1}}{(2n-1)!} \nonumber\\[2mm]
&=&\frac{1}{4\pi^2}\,\frac{(8\pi^2)^n}{(2n)!}\,.
\label{resultpn0}
\eea
This improved estimate for $p_n$ is
the end result of this section. The same expression
occurred at an intermediate stage of the calculation
in Ref.\,\cite{Hilhorst05b}, where it went by the name of $p_n^{(0)}$. 
Eq.\,(\ref{resultpn0}) 
has the $n$ dependence that we know is exact:
upon taking the logarithm we now recover the first three terms of 
expansion (\ref{resultlogpn}).
The exponential $(8\pi^2)^n$ and the factorial denominator
in (\ref{resultpn0}) are therefore correct.
However, our derivation has been too crude \cite{footnotec} 
that we may trust the 
proportionality constant $(4\pi^2)^{-1}$.
This constant turns out to
differ from the correct one appearing in (\ref{resultpn})
by the factor $C$.
In the next section we will briefly discuss the origin of this factor.

%%%%%%%%%%%%%%%%%%%%%%%%%%%%%%%%%%%%%%%%%%%%%%%%%%%%%%%%%%%%%%%%%%%%%%%%%%%%%
%%%%%%%%%%%%%%%%%%%%%%%%%%%%%%%%%%%%%%%%%%%%%%%%%%%%%%%%%%%%%%%%%%%%%%%%%%%%%

\section{Elastic deformations and Gabriel neighbors}
\label{secelastic}

In the preceding section we used that the cell perimeter
is close to a circle. 
However, since each of the first-neighbor
coordinates $\bP_m$ was integrated over a two-dimensional domain, 
the integral does not respect this circularity property exactly.
This section deals with the deviations from circularity. 
It will turn out that the quantitative effect of these deviations on the
sidedness $p_n$ is beyond 
the heuristic approach of this paper.

%%%%%%%%%%%%%%%%%%%%%%%%%%%%%%%%%%%%%%%%%%%%%%%%%%%%%%%%%%%%%%%%%%%%%%%%%%%%%

\subsection{Random acceleration process}
\label{secrap}

%%% [*** EXT 2002]
In this subsection $\bR_m$ will denote the position 
vector of the $m$th mid-point and $R_m\equiv |\bR_m|$ 
its length, for $m=1,2,\ldots,n$. The ratio between two
successive mid-point distances can be expressed as
$R_m/R_{m-1} = \cos\gamma_m/\cos\beta_m$\,, 
with the auxiliary angles $\beta_m$ and
$\gamma_m$ defined in Fig.\,\ref{figtransformation}. Hence
\beq
\frac{R_{m-1}R_{m+1}}{R_m^2} = 
\frac{\cos\beta_m\cos\gamma_{m+1}}{\cos\gamma_m\cos\beta_{m+1}}\,.
\label{ratioR}
\eeq
We set
\beq
R_m=R_{\rm c} + \delta R_m\,,
\label{defdeltaR} 
\eeq
where $R_{\rm c}=R^*$ is given by (\ref{exprRstar}), 
and expand both sides of Eq.\,(\ref{ratioR})
using that $\beta_m$, $\gamma_m$, and  $\delta R_m/R_{\rm c}$
are all small as $n$ gets large. 
To leading order this yields the second-order difference equation
\bea
\delta R_{m-1} -2\delta R_m + \delta R_{m+1} &=&
-\thalf R_{\rm c}\left(
\beta_m^2-\gamma_m^2+\gamma_{m+1}^2-\beta_{m+1}^2
\right)
\nonumber\\[2mm]
&=&-\thalf R_{\rm c}\left[
\xi_m(\beta_m-\gamma_m) - \xi_{m+1}(\beta_{m+1}-\gamma_{m+1})
\right], \nonumber\\
&&
\label{expr1DeltaR}
\eea
where we made use of the identity $\beta_m+\gamma_m=\xi_m$. 
In the second line of (\ref{expr1DeltaR}) 
the differences $\beta_\ell-\gamma_\ell$ (with $\ell=m,m+1$) are of zero
average. 
In that line
we now replace $\xi_m$ and $\xi_{m+1}$ by their average $2\pi/n$,
the idea being that the random contribution to these angles will have a
negligible effect when Eq.\,(\ref{expr1DeltaR}) is integrated 
(that is, summed on $m$) \cite{footnoted}. 
The right hand side of Eq.\,(\ref{expr1DeltaR}) having thus 
become linear in the angles, 
it appears that we can express the sum of the $\beta$'s and $\gamma$'s  
again in terms of the $\xi$'s and $\eta$'s.
In that way we obtain 
\beq
\delta R_{m-1} -2\delta R_m + \delta R_{m+1} =
\frac{\pi^\half}{(\rho n)^\half}\,F_m
\label{expr2DeltaR}
\eeq
with
\beq
F_m=\eta_m-\thalf(\xi_m+\xi_{m+1}).
\label{defFm}
\eeq
%%% [*** See I(6.12)]
The left hand side of Eq.\,(\ref{expr2DeltaR}) represents the ``radial
acceleration'' of the perimeter as it turns around the central seed.
The right hand side is a random term of known properties. 
An equation like (\ref{expr2DeltaR}) also appeared in
Refs.\,\cite{Hilhorst05b} and \cite{Hilhorstetal08}, where the conditions of
its validity were discussed. It was shown there 
that in the limit $n\to\infty$, with the replacement 
$m \mapsto \phi \equiv 2\pi mn^{-1}$,
the equation becomes a second-order differential equation 
$\dd^2\delta R(\phi)/\dd\phi^2=F(\phi)$ where $F(\phi)$ is Gaussian noise.
For white Gaussian noise this equation
is referred to as the {\it random acceleration process\,}
\cite{MajumdarComtet04,MajumdarComtet05,Gyorgyietal07}. 

For our present discussion Eq.\,(\ref{expr2DeltaR}) suffices as it stands. 
The key point is that $F_m$ and $F_{m'}$ are essentially 
uncorrelated when $|m-m'| \geq 2$. 
``Essentially'' here and below will mean: neglecting the weak
anticorrelation, of relative order ${\cal O}(n^{-1})$,
induced by the delta function constraint in the integral in (\ref{exprImidp}). 
The $F_m$ have zero mean and are, just like the $\xi_m$ and the $\eta_m$, 
of order $n^{-1}$ as $n$ gets large. 

Let us now sum both sides of (\ref{expr2DeltaR})
from $m=1$ to some value of $m$ which is of order $n$.
The right hand side becomes
a sum of ${\cal O}(n)$ zero-mean and essentially uncorrelated terms 
that are each of order $n^{-1}$; hence it will be of order $n^{-\half}$.
Taking also into account the
prefactor $(\pi/\rho n)^\half$ we find for the ``radial speed'' 
$\delta R_{m+1}-\delta R_m$ of the perimeter the expression 
\beq
(\delta R_{m+1}-\delta R_{m})\,-\,(\delta R_1-\delta R_0)\,=\,{\cal O}(n^{-1}).
\label{expr2dR}
\eeq
Since the average speed must be zero, this means that also
$\delta R_{m+1} - \delta R_m={\cal O}(n^{-1})$. 

\begin{figure}
\begin{center}
\scalebox{.45}
{\includegraphics{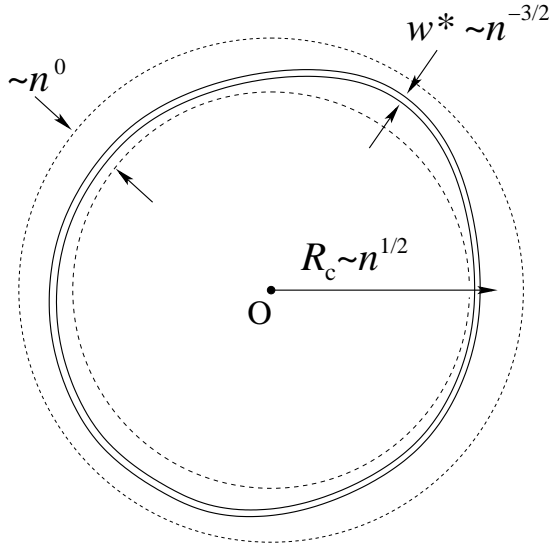}}
\end{center}
\caption{\small Pair of solid lines: the narrow annulus of width 
$w^*\sim n^{-\frac{3}{2}}$ through which the cell perimeter runs. Its 
radius has variations of order $n^0$ with respect to an 
average radius $R_{\rm c}\sim n^\half$ and stays, typically, between the two
concentric dashed circles.}
\label{figlargescale}
\end{figure}

Being the sum of essentially independent increments,
the radial speed $\delta R_{m+1} - \delta R_m$ varies with $m$ 
as a random walk trajectory.
Such a trajectory has strong long-range correlations.
Therefore, if we sum (\ref{expr2dR})
once more over ${\cal O}(n)$ values of the index $m$,
we obtain the sum of ${\cal O}(n)$ strongly
correlated terms that are each of order $n^{-1}$, which yields 
\beq
\delta R_m-\delta R_1={\cal O}(1).
\label{expr1dR}
\eeq
We have argued in section \ref{secestimating} 
that the cell perimeter is contained in an
annulus of width $\sim n^{-\frac{3}{2}}$.  
Eq.\,(\ref{expr1dR}) now shows that this annulus is not
strictly circular but that its radius deviates from the typical value 
$R_{\rm c}=(n/4\pi\rho)^\half$ by amounts of order unity.
In full coherence with this, an evaluation of the radial integral
(\ref{radialint}) by the steepest descent method yields
a peak width also of order unity.
The resulting large-scale behavior of the cell perimeter 
has been schematized in Fig.\,\ref{figlargescale}.
A ``real-life'' example of a 
not-quite-circular cell is provided by Fig.\,\ref{figcell96}. 

%%%%%%%%%%%%%%%%%%%%%%%%%%%%%%%%%%%%
%%%%%%%%%%%%%%%%%%%%%%%%%%%%%%%%%%%%
%%%%%%%%%%%%%%%%%%%%%%%%%%%%%%%%%%%%
%%%%%%%%%%%%%%%%%%%%%%%%%%%%%%%%%%%%

%%%%%%%%%%%%%%%%%%%%%%%%%%%%%%%%%%%%
%%%%%%%%%%%%%%%%%%%%%%%%%%%%%%%%%%%%
\begin{figure}
\begin{center}
\scalebox{.45}
{\includegraphics{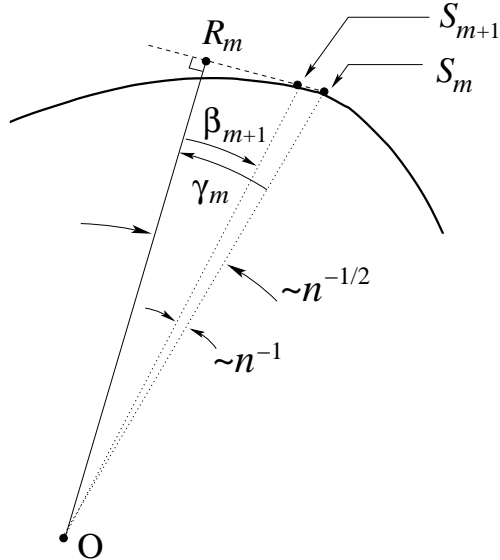}}
\end{center}
\caption{\small 
A cell perimeter (heavy solid line) for which
the mid-point $R_m$ does not lie on the $m$th
perimeter segment $S_mS_{m+1}$ but elsewhere on the line 
prolonging this segment. The angle $\beta_{m+1}$ is negative.
The $\gamma_m$ and $\beta_m$ are typically of order $n^{-\half}$ 
and the situation depicted in this figure is generic rather than 
exceptional.} 
\label{figbetagamma}
\end{figure}
%%%%%%%%%%%%%%%%%%%%%%%%%%%%%%%%%%%%
%%%%%%%%%%%%%%%%%%%%%%%%%%%%%%%%%%%%

%%%%%%%%%%%%%%%%%%%%%%%%%%%%%%%%%%%%%%%%%%%%%%%%%%%%%%%%%%%%%%%%%%%%%%%%%%%%%

\subsection{Origin of the constant $C$}
\label{secconstantC}

In section \ref{secrap} we identified the
random acceleration process as the mechanism that governs
the perimeter's deviations from circularity.
These deformations may be referred to as ``elastic,'' 
with the understanding that this elasticity is of purely
entropic origin. It is in fact a long-range, ``macroscopic''
elasticity that comes over and above the
short-range perimeter fluctuations already taken into account  in section
\ref{secimproved} by the integrations on the $\xi_m$ and $\eta_m$.

The exact determination of the effect on $p_n$ 
of these elastic deformations is beyond the
heuristic arguments of this paper. For completeness we recall that 
they may be treated \cite{Hilhorst05a,Hilhorst05b}
as a superposition of independent Fourier components
with wavenumbers $q=\pm 1,\pm 2,\ldots$.
Their amplitudes $\psi_q$ are Gaussian random variables and 
their elastic energy \cite{footnoteg}\,
$E=\pi \sum_{q}\Lambda_q\psi_{q}\psi_{-q}$\, 
has the dispersion relation
\beq
\Lambda_q=1-q^{-2}+4q^{-4}.
\label{exprLambdaq}
\eeq
The elastic deformations multiply $p_n$ by an extra contribution 
equal to their partition function $C$, which is 
obtained as $C=\prod_{q=1}^{\infty}\Lambda_q^{-1}$. 

%%%%%%%%%%%%%%%%%%%%%%%%%%%%%%%%%%%%%%%%%%%%%%%%%%%%%%%%%%%%%%%%%%%%%%%%%%%%%

\subsection{Gabriel neighbors of the many-sided cell}
\label{secGabriel}

In this subsection we apply the heuristic arguments developed
above to Gabriel neighborship.
Two adjacent Voronoi cells are called Gabriel (or full) neighbors 
if the line segment connecting their seeds does not intersect any other cell.
In the notation of this work, the condition for 
the $m$th first neighbor of the central seed to be a Gabriel
neighbor is that the mid-point $R_m$ lie on the 
perimeter segment $S_{m}S_{m+1}$. In the example of Fig.\,\ref{figbetagamma}
the $m$th neighbor is {\it not\,} Gabriel.
Reasons for being interested in Gabriel
neighbors have been mentioned in the Introduction. 

Whereas in two dimensions the average {\it total\,}
number of neighboring cells is exactly
$\la n\ra=6$, the average number of full neighbors is equal only
to $\la n\ra^{\rm full}=4$. 
We ask now: given an $n$-sided cell, what is its average number 
$\nu^{\rm full}_n$ of Gabriel neighbors?
The large-$n$ behavior of $\nu^{\rm full}_n$ can be estimated as follows.

Fig.\,\ref{figbetagamma} shows that $R_m$
is in the interval $S_mS_{m+1}$ if and only if $\gamma_m$ and
$\beta_{m+1}$ are both positive.
These conditions are not independent since  
$\gamma_m+\beta_{m+1}=\eta_{m+1}$ and $\eta_{m+1}$ is necessarily positive.
An equivalent condition for the $m$th first neighbor to be Gabriel
is that $\gamma_m>0$ and $\eta_m>\gamma_m$. 
The advantage of this formulation is that $\gamma_m$ and $\eta_m$ are
essentially independent (in the same sense as before); 
this is also clear from the algorithm described in section \ref{secimproved}
by which the perimeter is constructed on the basis of given sets of $\xi$'s
and $\eta$'s. 
Let us write $w_n(\gamma_m)$ and $v_n(\eta_m)$ for the probability 
distributions of the $\gamma_m$ and the $\eta_m$, respectively.
We then have
\beq
\nu^{\rm full}_n = n\int_0^\infty\!\dd\gamma\,w_n(\gamma) 
\int_\gamma^\infty\!\dd\eta\,v_n(\eta),
\label{exprnufulln}
\eeq
in which the prefactor $n$ comes from the total number of cell sides
and the remaining integral represents
the probability that $\gamma_m, \beta_{m+1}>0$.
We now need the distributions $v_n$ and $w_n$.
Eq.\,(\ref{exprIvert}) shows that the $\eta_m$ define the partition of
a circle into $n$ random and independent intervals of average 
length $2\pi/n$. Hence $v_n(\eta)$ is an exponential of average $2\pi/n$,
\beq
v_n(\eta)=(n/2\pi)^{-1}\exp\left( -n\eta/2\pi\right).
\label{exprvneta}
\eeq
In order to find $w_n$ we remark that any difference 
$\gamma_{m_2}-\gamma_{m_1}$ can be written as 
$\gamma_{m_2}-\gamma_{m_1}=\sum_{m=m_1+1}^{m_2}(\xi_m-\eta_{m-1})$,
which is a  sum of $2m_2-2m_1$ essentially independent random variables.
It follows that the differences $\gamma_{m_2}-\gamma_{m_1}$,
as well as the $\gamma_m$ themselves, are Gaussian distributed
with a root-mean-square width that scales as $n^{-\half}$.
Knowing that the $\gamma_m$ must have average $\pi/n$
and writing the asymptotic behavior of
their variance as $\simeq(c_0 n)^{-1}$ with $c_0$ a numerical constant, 
we have that in the large-$n$ limit
%%% [*** PVC 2616 sqq]
\beq
w_n(\gamma)=(c_0 n/2\pi)^{\half}
    \exp\left[ -\thalf c_0 n(\gamma-\pi/n)^2 \right].
\label{exprwngamma}
\eeq
Of course the angles $\beta_m$, related to the $\gamma_m$ by inversion of the
orientation of the perimeter, also have the distribution $w_n$.
We refer again to Fig.\,\ref{figbetagamma} for an additional remark.
Since $w_n$ is a distribution on the scale $n^{-\half}$ 
but the angles $S_mOS_{m+1}=\eta_m$ are of order $n^{-1}$,
the situation depicted in this figure
is generic: the perpendicular from the origin onto the line containing
a perimeter segment will typically hit that line in a point {\it outside\,} 
that segment.  

Substituting (\ref{exprvneta}) and (\ref{exprwngamma}) in (\ref{exprnufulln})
and evaluating in the limit of large $n$ we obtain the scaling with $n$
of the number of Gabriel neighbors,
\beq
\nu^{\rm full}_n \simeq (2\pi c_0 n)^\half, \qquad n\to\infty.
\label{resultnufulln}
\eeq
This confirms 
that for many-sided cells Gabriel neighbors are {\it atypical}.
The proportionality constant $c_0$ in (\ref{resultnufulln})
cannot be obtained by the present heuristic
arguments; it is, however, in principle accessible by a detailed
calculation in which, again, the $\Lambda_q$ will intervene.

%%%%%%%%%%%%%%%%%%%%%%%%%%%%%%%%%%%%%%%%%%%%%%%%%%%%%%%%%%%%%%%%%%%%%%%%%%%%%
%%%%%%%%%%%%%%%%%%%%%%%%%%%%%%%%%%%%%%%%%%%%%%%%%%%%%%%%%%%%%%%%%%%%%%%%%%%%%

\section{Conclusion}
\label{secconclusion}

We have developed heuristic arguments applicable to Voronoi cells
and whose validity is confirmed by existing exact calculations.
The arguments, which  concern Voronoi cells in the limit
of large sidedness,
are based on the estimation of the entropy of a set
of points arranged in a plane under specific geometrical constraints.
The power as well as the limitations of these arguments have been indicated.
As a new result we have determined by the reasoning of the present paper
the asymptotic $n$ dependence
of the average number of Gabriel neighbors of the $n$-sided cell.

We may henceforth apply these heuristic methods with a certain confidence 
in other contexts where no exact calculations are possible. 
Examples of related problems 
in mathematics and statistical physics 
have been  mentioned in the Introduction.
Finally, this work provides the tools that will be used 
in an upcoming study of Voronoi cells in 
higher spatial dimensions. 

%%%%%%%%%%%%%%%%%%%%%%%%%%%%%%%%%%%%%%%%%%%%%%%%%%%%%%%%%%%%%%%%%%%%%%%%%%%%%

\appendix


\begin{thebibliography}{10}
\bibitem{footnoteb}
Eqs.\,(\ref{resultpn}) and (\ref{resultlogpn}) are exact by the usual
standards of mathematical physics; 
however, they are still awaiting rigorous mathematical confirmation.
\bibitem{footnotee}
The accuracy of earlier Monte Carlo methods
\cite{DrouffeItzykson84,ItzyksonDrouffe89}
is down to the single-digit level
when $n\approx 20$.
\bibitem{footnoteg}
In analogy with many other problems in statistical physics, 
if $p$ is a phase space integral we will sometimes 
imagine it written as $p=\ee^S$ and call $S$ an entropy, or
as $p=\ee^{-E}$ and call $E$ an energy.
\bibitem{footnotec}
In fact, in (\ref{resultpn0}) a cancellation has occurred between 
two factors of distinct origin that we did not take into account 
in the argument of section \ref{secimproved}.
First, a trivial factor $2\pi$ comes from the rotational degree of freedom of
the cell. Secondly, a factor $(2\pi)^{-1}$ has a more subtle origin: 
it arises from an extra nonlinear constraint on the $\xi_m$ and $\eta_m$,
needed to guarantee that the perimeter closes onto 
itself after turning around the central seed.
\bibitem{footnoted}
Correlations between the $\beta_\ell$ and $\gamma_\ell$ on the one hand and
the $\xi_m$ and $\eta_m$ on the other, are negligible in the large-$n$ limit.
\end{thebibliography}
\end{document}